\documentclass[aps,prl,amsmath,twocolumn]
{revtex4}
\usepackage[dvips]{graphicx}
\newcommand{\PSbox}[3]{\mbox{\rule{0in}{#3}\includegraphics{#1}\hspace{#2}}}

\begin{document}
\draft

\title{Circular Polarization Dependent Study of the Microwave Photoconductivity in a Two-Dimensional Electron System}  \vskip -3.mm
\author{J.H.~Smet,$^{a}$ B.~Gorshunov,$^{a,d}$ C.~Jiang,$^{a}$ L.~Pfeiffer,$^{b}$
K.~West,$^{b}$ V.~Umanksy,$^{c}$ M.~Dressel,$^{d}$
R.~Meisels,$^{e}$ F.~Kuchar,$^{e}$ K.~von Klitzing,$^{a}$}

\affiliation{$^a$Max-Planck-Institute f\"ur
Festk\"orperforschung, Heisenbergstra\ss e 1, D-70569 Stuttgart, Germany\\
$^b$Lucent Technologies, Bell Labs, Murray Hill, NJ 07974, USA\\
$^c$Braun Center for Submicron Research, Weizmann Institute of
Science, Rehovot 76100, Israel \\
$^d$1. Physikalisches Institut, Universit\"at Stuttgart,
Stuttgart, Germany \\
$^e$Department of Physics, University of Leoben, A-8700 Leoben,
Austria}

\begin{abstract}
The polarization dependence of the low field microwave
photoconductivity and absorption of a two-dimensional electron
system has been investigated in a quasi-optical setup in which
linear and any circular polarization can be produced in-situ. The
microwave induced resistance oscillations and the zero resistance
regions are notedly immune to the sense of circular polarization.
This observation is discrepant with a number of proposed theories.
Deviations only occur near the cyclotron resonance absorption
where an unprecedented large resistance response is observed.

\vskip 5pt \noindent PACS numbers: 73.21.-b, 73.43.-f
\end{abstract}

\maketitle
 The recent discovery of zero
resistance induced by
microwaves~\cite{mani2002,zudov2003,dorozh2003,yang2003,mani2004,studenikin2004,willett2004}
in ultra-clean two-dimensional electron systems (2DES) over
extended regions of an applied perpendicular magnetic field $B$
has revived the general interest in microwave photoconductivity
and has triggered a remarkably large and diverse body of
theoretical works. Original photoconductivity experiments on lower
quality samples only revealed the intuitively expected feature due
to resonant heating at the cyclotron resonance or more accurately
--- due to the finite size of the sample --- at the combined
dimensional plasmon cyclotron resonance
frequency~\cite{vasiliadou1993}. Unanticipated $1/B$-periodic
oscillations with minima close to the harmonics of the cyclotron
resonance  first entered the scene~\cite{zudov2001,ye2001}. They
later turned out precursors of the zero resistance regions. The
majority of theoretical accounts subdivides the argumentation to
explain the zero resistance into two main points. First some
mechanism produces an oscillatory photoconductivity contribution
that may turn the overall dissipative conductivity negative near
the minima. Examples of proposed mechanisms include spatially
indirect inter-Landau-level transitions based on impurity and
phonon
scattering~\cite{ryzhii1970,ryzhii1986,durst2003,ryzhii2003,ryzhii2003_2,ryzhii2003_3,lei2003,shikin2003,vavilov2004},
the establishement of a non-equilibrium distribution
function~\cite{dorozh2003,dmitriev2004,dmitriev2005,dietel2005},
photon assisted quantum tunneling~\cite{shi2003} and
non-parabolicity effects~\cite{koulakov2003}. Second, it is argued
that negative values of the dissipative conductivity render the
initially homogeneous system
unstable~\cite{zakharov1960,andreev2003} and an inhomogeneous
domain structure develops
instead~\cite{andreev2003,bergeret2003,ryzhii2003_4,volkov2004,auerbach2005},
which results in zero resistance in experiment. Some theoretical
work does not invoke an instability driven formation of domains to
explain zero resistance. It relies either on radiation induced
gaps in the electronic spectrum~\cite{rivera2004} or on  the
microwave driven semi-classical dynamics of electron
orbits~\cite{inarrea2005}.

The sheer multitude of models and their divergence underline that
no consensus has been reached on the origin of this
non-equilibrium phenomenon. In order to assist in isolating the
proper microscopic picture, a detailed polarization dependent
study was carried out. In previous work, microwaves were guided to
the sample with oversized rectangular
waveguides~\cite{mani2002,zudov2003,dorozh2003,yang2003,mani2004}
or with a coaxial dipole
antenna~\cite{studenikin2004,willett2004}. These approaches do not
permit control over the polarization state. Rectangular waveguides
operated in their fundamental mode allow a comparison between the
two orthogonal linear polarization directions, although usually
not without warming up the sample. Here, we adopt a quasi-optical
approach~\cite{kozlov1998} to guide the microwaves onto the sample
and to produce any circular or linear polarization. Circular
polarization offers the perspective of activating and deactivating
the cyclotron resonance absorption by reversing the rotation for a
given $B$-field orientation. Knowledge of the influence of the
polarization on the microwave induced resistance oscillations may
turn out an important litmus test for theoretical models. For the
non-parabolicity model for instance these oscillations do not
survive when switching to circular
polarization~\cite{koulakov2003}. Theories based on a
non-equilibrium distribution function and on phonon- or impurity
assisted indirect transitions predict oscillations for both senses
of circular polarization, however with substantially different
amplitudes~\cite{dmitriev2005,vavilov2004}. Other models have
assumed linear polarization and have not been analyzed for the
case of circular
polarization~\cite{shi2003,rivera2004,inarrea2005}. Here, we
establish in experiment that these oscillations are entirely
insensitive to the polarization state of the incident radiation.

A $4 \times 4$-mm$^2$ van der Pauw sample is glued on a $25\ {\rm
\mu m}$ thick Mylar foil, which itself has a hole of 3 mm at the
location of the sample. This arrangement is mounted in the Faraday
geometry in the variable temperature insert of an optical cryostat
with a split coil. the insert was operated down to approximately
1.7-1.8 K during the experiments. The cryostat is equipped with
$100\ {\rm \mu m}$ thick Mylar inner and outer windows. The outer
windows are covered with black polyethylene foil to block visible
light. The sample of which we show data here consists of a
double-sided modulation doped 30 nm wide GaAs quantum well
surrounded by ${\rm Al_{0.24}Ga_{0.76}As}$ barriers. It exhibits
an electron mobility in excess of $18 \times 10^6\ {\rm cm^2}/{\rm
Vs}$ without prior illumination for a density of $2.6 \cdot
10^{11}/{\rm cm}^2$. Backward wave oscillators generate
quasi-monochromatic radiation (bandwidth $\Delta f/f \approx
10^{-5}$). Our studies focused on frequencies from 100 GHz up to
350 GHz.

\begin{figure}
\PSbox{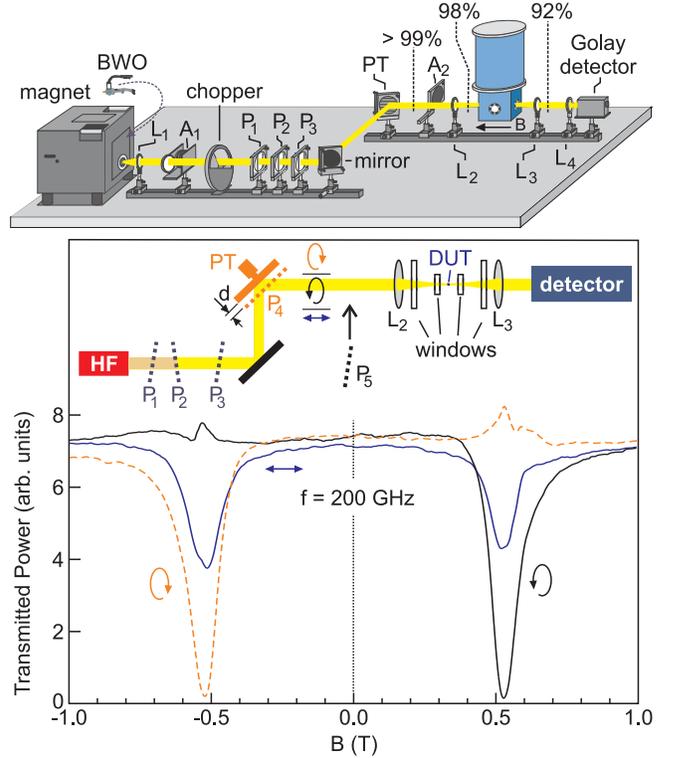}{8cm}{10cm} \caption{Top panel: Quasi-optical
setup for a polarization dependent study of the microwave
photoconductivity. Bottom panel: Transmission data for a frequency
of 200 GHz and linear, left or right hand circular polarized
radiation. The inset highlights those optical components
controlling or affecting the polarization state.} \label{setup}
\end{figure}

Schematic drawings of the quasi-optical setup~\cite{kozlov1998}
are depicted in Fig.~\ref{setup}. The radiation passes through
three dense wire grids with a periodicity of $50\ {\rm \mu m}$ and
a wire thickness of $20\ {\rm \mu m}$ ($P_{1}$, $P_{2}$, $P_{3}$)
as well as a so-called polarization transformer (PT). The latter
consists of the fixed wire grid $P_4$ with the same specifications
and a mobile metallic mirror placed in parallel to grid $P_4$ at a
tunable distance $d$. Grid $P_4$ reflects the component of the
incident radiation beam with the electric field vector aligned
along the wires. The remainder of the beam with an electric field
vector polarized perpendicular to the wires passes undisturbed
through the grid. It is reflected by the mirror instead and hence
acquires an additional phase shift $\Delta \phi = 2 \pi d
\sqrt{2}/\lambda$. Grids $P_1$, $P_2$ and $P_3$ serve to
continuously adjust the overall intensity  as well as to ensure
equal intensities of the radiation components with electric field
vectors aligned with and perpendicular to wire grid $P_4$, so that
proper adjustment of $d$ (or $\Delta \phi$) yields linear or any
circular polarization state. The radiation intensity can also be
reduced with fixed attenuator $\rm A_1$ or blocked with absorber
$\rm A_2$. The quartz lens $L_2$ focuses the radiation onto the
sample, while quartz lens $L_3$ recollimates the beam after
transmission through the cryostat. The degree of circular
polarization $\eta$ is verified at various locations along the
beam by monitoring the power with a pneumatic (Golay) detector
(used in conjunction with the chopper) and an additional dense
wire grid $P_5$ for two orthogonal directions ($||$ and $\perp$)
of the electric field vector: $\eta = 1 - | P_{\perp} - P_{||}
|/(P_{\perp} + P_{||})$. Typical values for the circular
polarization purity have been included in Fig.~\ref{setup}. After
the first quartz lens, but before entering the cryostat, $\eta$
exceeds $98\%$. The warped Mylar windows and the quartz lenses
deteriorate somewhat the polarization status, but the circular
polarization character remains better than $\eta
> 92 \%$ after transmission through the cryostat and sample holder
in the absence of a sample. Preliminary experiments, not shown
here, were also carried out in a setup with a circular microwave
waveguide and a polarizer at room temperature mounted at the
entrance of the waveguide. With this approach however the circular
polarization degree was limited to $70\%$ or less, presumably due
to imperfections of the stainless steel circular waveguide with a
diameter of 2.9 mm.

Fig.~\ref{setup} depicts the outcome of a transmission experiment
 for both circular polarization directions and linear
polarization of the incident radiation obtained with the
quasi-optical setup. Transmission experiments for unpolarized
radiation were reported previously in Ref.~\cite{studenikin2005}.
Data is shown for both positive and negative values of the
magnetic field. Active cyclotron resonance absorption should only
occur for the proper sense of circular polarization with respect
to the magnetic field orientation. Indeed, the transmitted power
drops nearly to zero for cyclotron resonance absorption in the
active sense of the polarization (CRA), whereas reversing the
magnetic field orientation while maintaining the circular
polarization direction turns the cyclotron resonance mode inactive
(CRI). For linear polarization, the transmitted power does not
drop below $50\%$ as it should. These transmission data confirm
the quality of the various polarization states. We note that as in
Ref.~\cite{vasiliadou1993} the resonance should not occur at the
bare cyclotron frequency, but rather at the frequency of the
combined dimensional plasmon cyclotron resonance mode:
$(\omega_{\rm c}^2 + \omega_{\rm dp}^2)^{1/2}$. Here, $\omega_{\rm
dp}$ is the plasmon frequency at $k=\pi/L$ with $L$ the size of
the sample. The estimated value of $\omega_{\rm dp}$ is 12 GHz.
Hence,
 at a microwave frequency of 200 GHz, the combined mode is only
0.5 \% larger than the bare cyclotron frequency. In view of the
large conductivity of the 2DES, the radiation is mainly reflected
near the resonance for the active circular polarization sense and
the linewidth is not determined by the scattering rate. It is
broadened orders of magnitude and this broadening has been
referred to as saturation effect~\cite{heitmann1986} or radiative
decay~\cite{mikhailov2004}. Noteworthy is also the absence of
discernible absorption signals at the harmonics of the cyclotron
resonance frequency. The dissipative photoresistivity is far more
sensitive and exhibits microwave induced oscillations up to the
10th harmonic of the cyclotron resonance (see data below).

\begin{figure}[b!] \PSbox{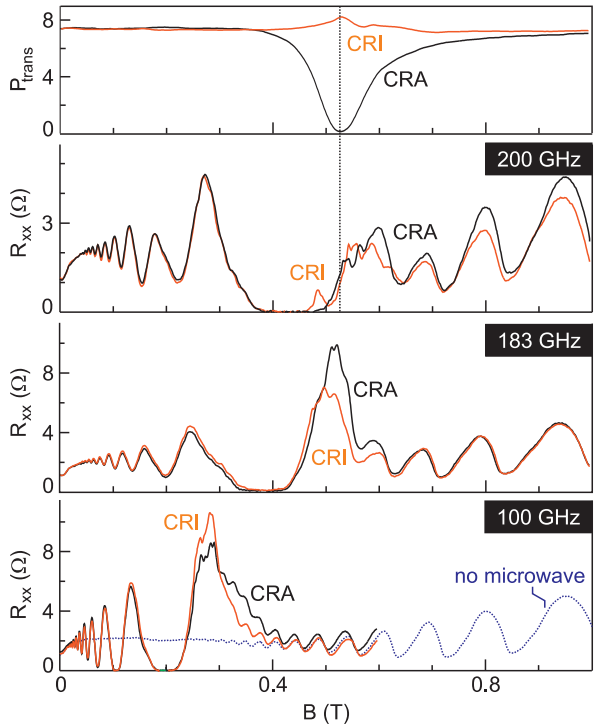}{8cm}{10cm}
\caption{Comparison of the magnetoresistivity under microwave
radiation for both senses of circular polarization and three
different frequencies. The transmitted power $P_{\rm trans}$ (arb.
units) at a frequency of 200 GHz is plotted at the top for the
sake of comparison with the 200 GHz transport data. The blue
dotted curve in the bottom panel displays the magnetoresistivity
in the absence of microwave radiation.} \label{comparisonZRS}
\end{figure}

Fig.~\ref{comparisonZRS} illustrates the influence of the circular
polarization sense on the microwave induced resistance
oscillations for different microwave frequencies: 200, 183 and 100
GHz. The absorption signals for 200 GHz have been included at the
top to allow for easy comparison of any features relative to the
cyclotron resonance position. At fields below the cyclotron
resonance absorption where the microwave induced oscillations
occur, the magnetoresistivity traces for both senses of circular
polarization are nearly indistinguishable. The same conclusion
holds for linearly polarized radiation irrespective of the
orientation of the electric field vector (an example is shown
below in Fig.~\ref{spike}). To rule out that the microwave induced
oscillatory photoresistivity has saturated, data were taken at
power levels three times as high as for the traces plotted in
Fig.~\ref{comparisonZRS}. The amplitude of the second and higher
order oscillations increased by as much as a factor of two. So we
can safely discard the possibility that the presence of the small
residual power of radiation with the undesirable polarization
direction in conjunction with saturation of the microwave induced
oscillations would produce this apparent polarization
insensitivity. Hence, the experiments here only support
theoretical mechanisms in which the polarization state of the
microwaves is not relevant. Only at fields near the cyclotron
resonance, where significant absorption takes place, deviations
between the CRA and CRI curves do arise (see also below). The
experimental observation that the higher order maxima and minima
remain unaltered when reversing the sense of circular polarization
may be regarded as a crucial test for theories.  For instance, it
contradicts the non-parabolicity model which only produces an
oscillatory photoconductivity in the case of linear
polarization~\cite{koulakov2003}. It can also not be reconciled
with the two most frequently cited theories, for which the
influence of the circular polarization sense can easily be
analyzed with analytical formulas reported in the
literature~\cite{dmitriev2005}: the non-equilibrium distribution
function
scenario~\cite{dorozh2003,dmitriev2004,dmitriev2005,dietel2005}
and the picture based on impurity- and phonon assisted indirect
inter-Landau level
transitions~\cite{ryzhii1970,ryzhii1986,durst2003,ryzhii2003,ryzhii2003_2,ryzhii2003_3,lei2003,shikin2003,vavilov2004}.
As an example, in the linear response regime and for a not too
strong microwave field, the correction to the dark dc dissipative
conductivity is a factor $(\omega - \omega_{\rm c})^2/(\omega +
\omega_{\rm c})^2$ smaller for the CRI polarization sense (Eq. 16
and 17 in Ref.~\cite{dmitriev2005} with $\cal P_{\omega}$ adapted
for circular polarization. See also Eq. 6.11 in
Ref.~\cite{vavilov2004}). This factor originates in essence from
the difference in the ac Drude conductivity for both circular
polarization directions. For the second and third maximum (or
minimum) this amounts to a factor of about 9 and 4 respectively!
For larger microwave fields these corrections no longer obey a
linear but rather a sublinear dependence on the microwave power
and these factors are expected to be somewhat reduced. The
dramatic disparity with the experiment of Fig.~\ref{comparisonZRS}
however remains. There clearly is a strong need to analyze the
polarization dependence of other proposed theoretical mechanisms.

The primary maximum does depend on the sense of circular
polarization, likely because resonant heating at the cyclotron
resonance produces a second contribution to the photoresistivity.
In much older microwave photoconductivity experiments prior to the
discovery of the zero resistance phenomenon, the cyclotron
resonance absorption was already detected as a small resistance
increase~\cite{vasiliadou1993}. In the context of the recently
discovered microwave induced resistance oscillations, signatures
of the cyclotron resonance absorption in the magnetoresistivity
remained unidentified, presumably because they were masked by the
much larger microwave induced resistance oscillations. The
insensitivity of these oscillations on the polarization state and
the ability to in-situ alter the circular polarization direction
make it straightforward to reveal this bolometric contribution to
the photoresistivity. More striking examples appearing at high
power are illustrated in Fig.~\ref{spike} for frequencies of 243
and 254 GHz (data at 183 and 200 GHz at three times larger power
than in Fig.~\ref{comparisonZRS} also show a similar feature). The
resistance enhancement can be surprisingly large and develops fine
structure at these power levels. Its close connection to the
cyclotron resonance is established by comparing with transmission
data. The cyclotron resonance lineshape is distorted as a result
of standing waves in the sample as well as in windows and other
optical components as can be verified by investigating
transmission data for a large frequency range. The power
dependence of one such resistance peak is depicted in the bottom
insert of Fig.~\ref{spike}. We interprete the treshold like
behavior as a confirmation for the existence of two contributions
to the resistance peak: the bolometric resistance enhancement with
a nearly linear but steep power dependence and the contribution
related to the oscillatory photoresistivity.
\begin{figure}
\PSbox{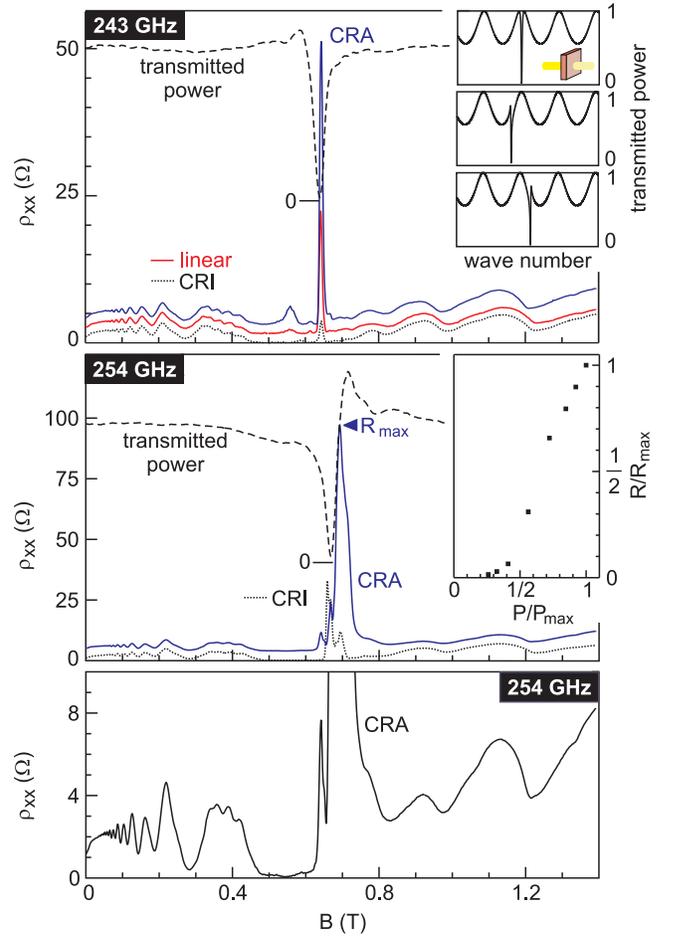}{8cm}{12.4cm} \caption{Top: The
magnetoresistivity for various polarizations under 243 GHz and 254
GHz radiation. Curves are offset for clarity. The dashed line
represents the transmitted microwave power (no ordinate shown).
The bottom panel is a blown up version of the 254 GHz data. The
top inset schematically illustrates the influence of the
unavoidable standing waves within the substrate on the lineshape
of the cyclotron resonance line. The bottom insert plots the
resistance maximum marked by a blue triangle as a function of the
microwave power. The maximum incident power and the corresponding
peak height are denoted as $P_{\rm max}$ and $R_{\rm max}$.}
\label{spike}
\end{figure}

In summary, investigations in an all quasi-optical setup offering
full control over the polarization properties of microwave
radiation disclosed in discord with common theoretical pictures a
complete immunity of the microwave induced resistance oscillations
to the polarization state. Only the bolometric signal associated
with cyclotron resonance absorption itself is strongly
polarization dependent. This puzzling discrepancy between theory
and experiment will likely reinvigorate the controversy on the
origin of the microwave induced oscillations.

We gratefully acknowledge discussions with I. Dmitriev, A. Mirlin
and D. Heitmann as well as support from the German Israeli
Foundation and the German Physical Society.

\end{document}